\documentclass[journal]{IEEEtran}

\usepackage{subfig}
\usepackage{caption}
\usepackage{float}
\usepackage{dblfloatfix} 
\usepackage{setspace}
\usepackage{graphicx}
\usepackage{graphicx,dblfloatfix}
\usepackage{tikz}
\usepackage{color}
\usepackage{amsmath}
\usepackage{amssymb} 

\usepackage{maplestd2e}
\usetikzlibrary{shapes,arrows,shadows}
\usepackage{xcolor}
\usepackage{lipsum}
\usepackage{cite,bm}
\allowdisplaybreaks

\def\Nw{N_{\mathrm{w}}}
\def\nt{n_{\mathrm{t}}}
\def\Nt{N_{\mathrm{T}}}

\def\j{\mathrm{j}}

\begin{document}

\title{Deep Learning-Aided Perturbation Model-Based Fiber Nonlinearity Compensation}

\author{Shenghang~Luo,~\IEEEmembership{Student Member,~IEEE}, Sunish~Kumar~Orappanpara~Soman,~\IEEEmembership{Senior Member,~IEEE,} Lutz~Lampe,~\IEEEmembership{Senior Member,~IEEE}, and  Jeebak~Mitra,~\IEEEmembership{Member,~IEEE}

	\thanks{Shenghang Luo and Lutz Lampe are with the Department of Electrical and Computer Engineering, The University of British Columbia, Vancouver, Canada (e-mail: $\{$shenghang,lampe$\}$@ece.ubc.ca).}
	\thanks{Sunish Kumar Orappanpara Soman was with the Department of Electrical and Computer Engineering, The University of British Columbia, Vancouver, Canada. He is now with the School of Engineering, Ulster University, Belfast, United Kingdom (e-mail: S.Orappanpara\_Soman@ulster.ac.uk).}
	\thanks{Jeebak Mitra was with Huawei Technologies, Ottawa, Canada. He is now with Dell Technologies Canada (e-mail: jeebak.mitra@dell.com).}
 \thanks{This work was supported by  Huawei Tech., Canada. The research was supported in part through computational resources and services provided by the Digital Research Alliance of Canada
(www.alliancecan.ca) and the Advanced Research Computing at the University of British Columbia.}}

\maketitle

\begin{abstract}
Fiber nonlinearity effects cap achievable rates and ranges in long-haul optical fiber communication links. Conventional nonlinearity compensation methods, such as perturbation theory-based nonlinearity compensation (PB-NLC), attempt to compensate for the nonlinearity by approximating analytical solutions to the signal propagation over optical fibers. However, their practical usability is limited by model mismatch and the immense computational complexity associated with the analytical computation of perturbation triplets and the nonlinearity distortion field. Recently, machine learning techniques have been used to optimise parameters of PB-based approaches, which traditionally have been determined analytically from physical models. It has been claimed in the literature that the learned PB-NLC approaches have improved performance and/or reduced computational complexity over their non-learned counterparts. In this paper, we first revisit the acclaimed benefits of the learned PB-NLC approaches by carefully carrying out a comprehensive performance-complexity analysis utilizing state-of-the-art complexity reduction methods. Interestingly, our results show that least squares-based PB-NLC with clustering quantization has the best performance-complexity trade-off among the learned PB-NLC approaches. Second, we advance the state-of-the-art of learned PB-NLC by proposing and designing a \textit{fully learned} structure by adopting the noiseless Manakov equation as the channel propagation model. We apply a bi-directional recurrent neural network for learning perturbation triplets that are alike those obtained from the analytical computation and are used as input features for the neural network to estimate the nonlinearity distortion field. Finally, we demonstrate through numerical simulations that our proposed fully learned approach achieves an improved performance-complexity trade-off compared to the existing learned and non-learned PB-NLC techniques.  

\end{abstract}

\begin{IEEEkeywords}
Optical fiber communication, nonlinearity compensation, machine learning, perturbation theory-based nonlinearity compensation, recurrent neural network.
\end{IEEEkeywords}

\IEEEpeerreviewmaketitle

\section{Introduction}

\IEEEPARstart{I}{n} recent years, the communication traffic through the core network  has increased at an unprecedented scale. This has mainly been fueled by the proliferation of digital applications and services, for example, e-learning, digital health, and e-agriculture \cite{ASharma2021, DBertsimas2021, DChen2021}. The continuing growth of such applications and services has been the cause for the ``capacity crunch" problem and imposes a strong demand to further expand the core network infrastructure  \cite{SHu2021,DRafique2016, PJWinzer2017, RJEssiambr2010}. 

To avoid network capacity shortages, polarization-division multiplexed (PDM) long-haul optical fiber communication systems, which are key constituents of the core communication network, have been forced to shift their operating region into the nonlinear regime of the optical fiber. In this regime, the optical intensity-dependent electro-optical effect, referred to as the Kerr effect, limits the transmission capacity \cite{Amari2017}. Therefore, the past decade has witnessed increased research efforts towards the design of digital signal processing (DSP)-based fiber nonlinearity compensation (NLC) techniques, suitable to deal with the detrimental effects of fiber nonlinearity \cite{Amari2017, Dar2017, SunishKumar2017, Rafique2016, Vassilieva2019, Liang2014, TTNguyen, OSSunishKumar2019}. Digital back-propagation (DBP) is a well-investigated DSP technique to compensate for fiber nonlinearity \cite{Ip2008, JCCartledge2017}. The DBP technique is based on the numerical solution of the Manakov equation using the split-step Fourier method (SSFM) with negated channel parameters \cite{Ip2008}. However, the DBP technique is not practically viable due to the computational load associated with the SSFM algorithm \cite{SKOSoman2021}. 

In contrast to the numerical approach of DBP, the solution of the Manakov equation can also be analytically approximated using regular perturbation theory \cite{Vannucci2002, VOliari2020, ARedyuk2020, YGao2014, LDou2012, AMecozzi2000}. The perturbation theory-based recursive closed-form approximate solution is in a power series form, and it converges to the exact solution asymptotically \cite{VOliari2020}. Intending to design a computationally efficient digital NLC technique, the perturbation series truncated to first-order (FO) has been well investigated in the literature for the post-compensation of the nonlinearity effect \cite{AMecozzi2000}. This technique was introduced in \cite{Tao2011} and is referred to as perturbation theory-based NLC (PB-NLC). The conventional (CONV) PB-NLC technique relies upon various simplifying assumptions, such as the single-span assumption for the whole link, lossless assumption, and Gaussian shape assumption for the input pulse shape \cite{AMecozzi2000, Tao2011}. While these simplifying assumptions make PB-NLC computationally more efficient than the DBP technique \cite{Tao2011}, they result in a reduced NLC performance when applied to realistic links \cite{OSSKumarSPPCom2021, JOSunish2021}. A few works in the literature attempted to break one or more of those assumptions to improve the PB-NLC performance, albeit with an increase in computational complexity. For example, the authors in \cite{YZhao2013ECOC}, and \cite{Tao2015JLT} include the power profile in the computation of the perturbation coefficients, and the work in \cite{FFrey2018ECOC} uses a realistic pulse shape in the computation of the nonlinearity distortion field. In \cite{Liang2014}, the authors proposed an additive-multiplicative (AM) PB-NLC technique to solve the approximation error caused by the power overestimation problem of the CONV PB-NLC technique resulting from the truncation of the PB series at first-order.  

Recently, machine learning (ML)-based approaches have extensively been investigated in optical communication systems \cite{RMButler2020, SZhang2019, YGao2019, CHager2018, Fougstedt2018, CHager2020, OSSKumarSPPCom2021}. The ML-based solutions effectively replace the CONV DSP techniques for signal impairment compensation. Furthermore, rather than relying on traditional ML techniques, some designs apply learning to existing structures used in DSP-based NLC. This approach is referred to as model-based ML \cite{RMButler2020}. For example, a fully connected neural network (NN) with the FO intra-channel cross-phase modulation (IXPM) and intra-channel four-wave mixing (IFWM) perturbation triplets as the input features has been proposed in \cite{SZhang2019} for fiber NLC.  This model-based approach, which we refer to as feedforward NN-assisted PB-NLC (FNN PB-NLC), utilizes the NN to effectively learn the physical dynamics of the nonlinear interaction in the optical fiber \cite{SZhang2019}. It effectively overcomes the limitations imposed by the various simplifying assumptions underlying the CONV PB-NLC technique. Extending on the FNN PB-NLC technique, \cite{YZhao2020Access} proposed using a recurrent neural network (RNN) to process the triplets, which achieves similar performance as FNN PB-NLC with about 47\% lower computational complexity. A convolution neural network (CNN) has also been investigated to process the triplets, as shown in \cite{Lietal}. Alternative to the NN-based PB-NLC techniques, the use of the least square (LS) method to estimate the perturbation coefficients, which we refer to as LS PB-NLC, has been proposed in  \cite{Malekiha2016, ARedyuk2020}. 
 
 It is important to note that in high baud-rate long-haul transmission links, the overall computational complexity of the mentioned NN-based NLC techniques may exceed that of the benchmark DBP technique. This is because the computational complexity is dominated by the triplet computation for a given symbol window length \cite{YZhao2020Access}. Hence,  one direction for the learned PB-NLC techniques has been to lower the complexity by pruning the unimportant weights of the feature processing neural network \cite{SZhang2019} and by using a cyclic buffer (CB) in the triplet computation stage \cite{ARedyuk2020}. Similarly, K-means clustering to quantize PB coefficients has been applied to CONV PB-NLC and LS PB-NLC methods \cite{Malekiha2016, ARedyuk2020}.

While the developed learned solutions have significantly expanded the family of PB-NLC methods, their relative improvements have often been shown by considering only a single benchmark scheme and by ignoring complexity reduction methods mentioned above. This motivates us, in this paper, to conduct a more comprehensive comparison of the complexity-performance trade-offs between the learned and non-learned PB-NLC approaches. 
 We have already presented initial results of this work in the conference version \cite{OurECOC}, and we include and discuss those results here for completeness. In addition to the complexity-performance comparison, we propose a new fully learned (FL) NLC scheme. In FL NLC, we use a bidirectional-long short-term memory (bi-LSTM)/bidirectional-gated recurrent unit (bi-GRU) network to estimate the triplet-like input features and an FNN for the feature processing to estimate the  nonlinear  distortion field. 

 
 The main contributions of this paper are summarized as follows:
\begin{itemize}

		
	
	

  
 	
 	\item We perform a comprehensive computational complexity-performance comparison for the existing learned and non-learned PB-NLC techniques by applying the available complexity reduction methods, aiming to present a precise trade-off analysis between them. In this part of the work, we also propose a learned version of the AM PB-NLC technique, referred to as the FNN-AM PB-NLC and include it in the trade-off comparison.  
 	
 	\item We propose and design FL NLC, which is a new fully learned NLC approach that learns  triplet-like input features using a bi-LSTM/bi-GRU network and estimates the nonlinearity distortion field using an FNN. 
 	
 	\item We include the FL NLC technique in the complexity-performance trade-off comparison with the existing PB-NLC methods. The results show that our proposed FL NLC outperforms all other learned and non-learned PB-NLC techniques.   
 	
\end{itemize}

The remainder of this paper is organized as follows. Section~\ref{s:system} describes the transmission system and optical fiber channel model. In Section~\ref{s:convpbnlc}, we briefly review perturbation theory and the AM perturbation model. Section~\ref{s:learned}  introduces learned PB-NLC methods. After summarizing existing methods, we present our  FNN-AM PB-NLC extension and the new FL NLC method. Section~\ref{s:complexity} provides an evaluation of computational complexity and complexity reduction methods. Following that, Section~\ref{s:results} presents the performance-complexity comparisons for the different methods. Finally, we conclude the paper in Section~\ref{s:conclusions}. 

\section{System Model}
\label{s:system}

\begin{figure*}[!t]
	\centering{}\includegraphics[width=0.90\textwidth,height=0.16\paperheight]{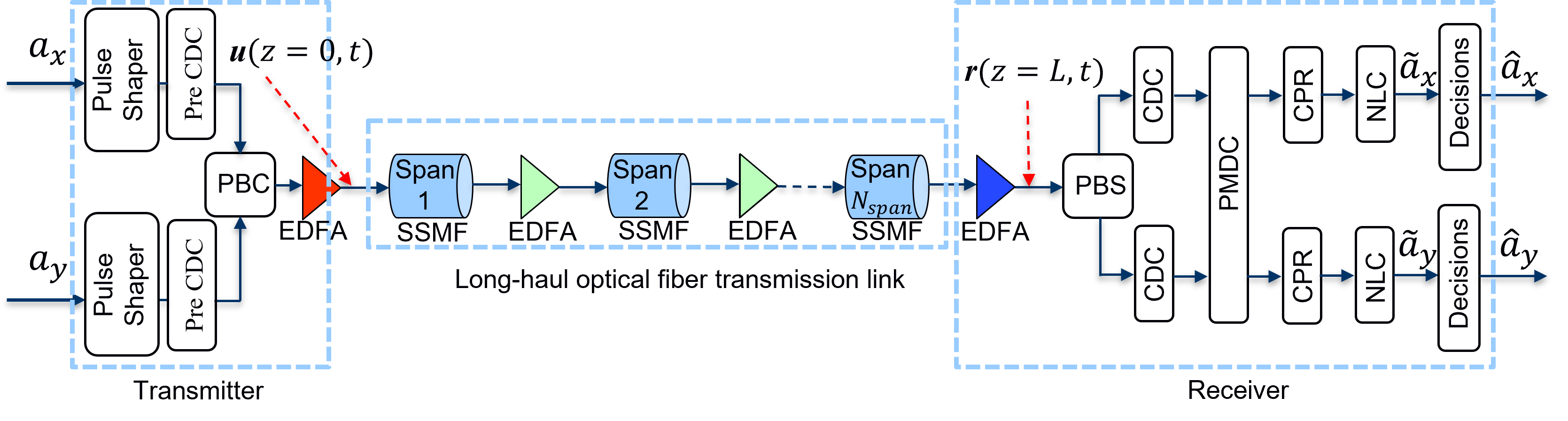}\caption{{The system model for the PDM system. PBC: polarization beam combiner, SSMF: standard single-mode fiber, EDFA: erbium-doped fiber amplifier, PBS: polarization beam splitter, CDC: chromatic dispersion compensation. PMDC: polarization-mode dispersion compensation: CPR: carrier phase recovery.}}
\label{fig1}
\end{figure*}

\subsection{Transmission System}

The PDM system model considered in this work is shown in Fig.~\ref{fig1}. At the transmitter, a sequence of $K$ symbols $\bm{a}_{x/y}=\left[a_{1,x/y},a_{2,x/y},\cdots,a_{K,x/y}\right]\in\Omega^{K}$, with $x,$ $y$ as the horizontal and vertical polarizations, respectively and $\Omega$ as the symbol alphabet, are passed through a pulse shaping filter $g(t)$, where $t$ represents the time. Following that, the signal is pre-compensated for 50\% of the chromatic dispersion (CD) to reduce the symbol interaction. The signal is then passed through a polarization beam combiner (PBC) to generate the PDM signal. The PDM signal is amplified to a given average power $P$ using a booster erbium-doped fiber amplifier (EDFA). The amplified signal envelop $\bm{u}(z=0,t)=\sqrt{P}\sum_{k}\bm{a}_{k,x/y}g(t-kT),$ where $T$ is the symbol duration, is then transmitted through the long-haul optical fiber transmission link. At the receiver, the signal field $\bm{r}(z=L,t),$ with $L$ as the transmission length, is polarization demultiplexed and converted into baseband. The resultant signal is then compensated for the remaining CD and the  polarization-mode dispersion (PMD) and passed through the carrier phase recovery (CPR) and fiber NLC blocks. Finally, the compensated symbol $\tilde{a}_{x/y}$ is input to the decision block to generate an estimate of the transmitted symbol, represented as $\hat{a}_{x/y}$.   

\subsection{Optical Fiber Propagation Model}

The Manakov equation is a well-known nonlinear partial differential equation to model the PDM signal propagation in the optical fiber channel \cite{SunishThesis}. In this work, we adopt a noiseless version of the Manakov equation as the channel propagation model, which at propagation distance $z$ is given as \cite{SunishThesis}
\begin{multline}
	\label{eqn1}
	\frac{\partial}{\partial z}\bm{u}(z,t)+\j\frac{\beta_{2}}{2}\frac{\partial^{2}}{\partial t^{2}}\bm{u}(z,t)\\
	=\j\frac{8}{9}\gamma\left|\bm{u}(z,t)\right|^{2}\bm{u}(z,t)\exp\left(-\alpha z\right),
\end{multline}
where 
 $\j$ is the imaginary unit, $\alpha$ is the attenuation coefficient, $\gamma$ is the nonlinearity coefficient, and $\beta_{2}$ is the group velocity dispersion coefficient. 
We solve the Manakov equation numerically using SSFM to model the evolution of the PDM signal envelop inside the optical fiber transmission link.       

\section{Conventional PB Theory and AM-PB-NLC} 
\label{s:convpbnlc}
The solution of the Manakov equation can be analytically approximated using perturbation theory. In the perturbation approach, the first-order nonlinear distortion added to a symbol at an arbitrary symbol index $i$ can be represented as a nonlinear beating between the symbols at the symbol indices $i+m$, $i+n$, and $i+m+n$. The corresponding expression for distortion is \cite{SunishThesis}
\begin{multline}
	\label{eqn2}
	d_{i,x/y}=\j P^{3/2}\sum_{m=-\Nw}^{\Nw}\sum_{n=-\Nw}^{\Nw}\left(a_{i+m,x/y}a_{i+m+n,x/y}^{*}\right.\\
	\left.+a_{i+m,y/x}a_{i+m+n,y/x}^{*}\right)a_{i+n,x/y}C_{m,n},
\end{multline}
where 
$\Nw$ is the symbol window length to calculate the nonlinearity distortion field. In (\ref{eqn2}), the perturbation triplets can be defined as
\begin{multline}
\label{e:triplets}
F_{i, (m,n), x/y}=(a_{i+m,x/y}a_{i+m+n,x/y}^{*}+\\
a_{i+m,y/x}a_{i+m+n,y/x}^{*})a_{i+n,x/y},
\end{multline}
and we denote the total number of triplets as $\nt$. Similarly, in (\ref{eqn2}), $C_{m,n}$ represent the perturbation coefficients, which can be represented as \cite{Liang2014}
\begin{multline}
	\label{eqn3}
	C_{m,n}=\frac{1}{T}\intop_{0}^{L}dz\exp(-\alpha z)
	\intop_{-\infty}^{\infty}dtg^{*}(z,t)\\ g(z,t-mT)g(z,t-nT)g^{*}(z,t-(m+n)T),
\end{multline}  
where 
 $g(z, t)$ is the pulse waveform that evolves spatially and temporally along the optical fiber. 

The perturbative distortion field in (\ref{eqn2}) consists of self-phase modulation (SPM), IXPM, and IFWM effects. The perturbation analysis generally models the nonlinear signal propagation in the optical fiber as an additive nonlinearity model, where the first-order distortion field is added to the transmitted symbol. For example, for symbol index $i=0$, the modified data symbol because of nonlinear distortion can be written as 
\begin{equation}
a'_{0,x/y}=a_{0,x/y}+\gamma d_{0,x/y}.
\end{equation}
Then, after applying the PB-NLC at the receiver, the estimated transmitted symbol follows as
\begin{equation}
	\label{eqnPBResult}
	\tilde{a}_{0,x/y}=r_{0,x/y}-\Delta\mu_{\textrm{add},0,x/y},
\end{equation}
where $r_{0,x/y}$ is the complex sample at $i=0$ after CD and PMD compensation, 
$\Delta\mu_{\textrm{add},0,x/y}=\gamma d'_{0,x/y}$ with $d'_{0,x/y}$ is the nonlinearity distortion field obtained by approximating transmitted symbols $a_{0,x/y}$ by $r_{0,x/y}$ in (\ref{eqn2}). In order to effectively model the phase noise characteristics of the SPM and a part of the IXPM termed as coherent IXPM (CIXPM), one can adopt an AM model \cite{Liang2014}. 
 This can be illustrated by rewriting (\ref{eqn2}) as shown in (\ref{eqn7}).
\begin{figure*}[!t]
	\begin{multline}
		\label{eqn7}
d_{0,x/y}=
\j P^{3/2}\underset{\textrm{SPM+CIXPM}}{\underbrace{a_{0,x/y}\left(\left(\left|a_{0,x/y}\right|^{2}+\left|a_{0,y/x}\right|^{2}\right)C_{0,0}+\sum_{m\neq0}\left(2\left|a_{m,x/y}\right|^{2}+\left|a_{m,y/x}\right|^{2}\right)C_{m,0}\right)}}\\
		+\j P^{3/2}\underset{\textrm{ICIXPM}}{\underbrace{\sum_{m\neq0}a_{0,y/x}a_{m,y/x}^{*}a_{m,x/y}C_{m,0}}}+\j P^{3/2}\underset{\textrm{IFWM}}{\underbrace{\sum_{m\neq0}\sum_{n\neq0}\left(a_{m,x/y}a_{m+n,x/y}^{*}+a_{m,y/x}a_{m+n,y/x}^{*}\right)a_{n,x/y}C_{m,n}}}
	\end{multline}
	\rule{1\textwidth}{0.5pt}
\end{figure*}
In the AM model, the SPM and CIXPM are modelled as multiplicative terms and the remaining part of IXPM, termed as incoherent IXPM (ICIXPM), and the IFWM terms are represented as additive terms. For this, we start with the definitions
\begin{multline}
	\label{eqn8}
	\phi_{\textrm{NL},0,x/y}=\gamma P^{3/2}\left(\left(\left |a_{0,x/y}\right |^{2}+\left |a_{0,y/x}\right |^{2}\right)C_{0,0}\right.\\
	\left.+\sum_{m\neq 0}\left(2\left |a_{m,x/y}\right |^{2}+\left |a_{m,y/x}\right |^{2}\right)C_{m,0}\right),
\end{multline}
\begin{equation}
	\label{eqn9}
	\Delta a_{\textrm{ICIXPM},0,x/y}=\j\gamma P^{3/2}\sum_{m\neq 0}a_{0,y/x}a_{m,y/x}^{*}a_{m,x/y}C_{m,0},
\end{equation}
and 
\begin{multline}
	\label{eqn10}
	\Delta a_{\textrm{IFWM},0,x/y}=\j\gamma P^{3/2}\sum_{m\neq0}\sum_{n\neq0}\left(a_{m,x/y}a_{m+n,x/y}^{*}\right.\\
	\left.+a_{m,y/x}a_{m+n,y/x}^{*}\right)a_{n,x/y}C_{m,n}.
\end{multline}
Then, we can represent the modified data symbol at time index $i=0$ as 
\begin{align}
	\label{eqn11}
a'_{0,x/y} & =a_{0,x/y}+\gamma d_{0,x/y}\nonumber \\
 & =a_{0,x/y}\left(1+j\phi_{\textrm{NL},0,x/y}\right)+\Delta a_{\textrm{ICIXPM},0,x/y}\nonumber \\
 & \hspace{3cm}+\Delta a_{\textrm{IFWM},0,x/y}\nonumber \\
 & \thickapprox a_{0,x/y}\exp\left(j\phi_{\textrm{NL},0,x/y}\right)+\Delta a_{\textrm{ICIXPM},0,x/y}\nonumber \\
 & \hspace{3cm}+\Delta a_{\textrm{IFWM},0,x/y}.
\end{align}

Equation (\ref{eqn11}) represents the AM model for the nonlinear signal propagation in the optical fiber. The NLC technique using this model is referred to as AM PB-NLC. The estimated data is given by \cite{Liang2014}
\begin{equation}
	\label{eqn12}
	\tilde{a}_{0,x/y}=(r_{0,x/y}-\Delta\mu'_{\textrm{add},0,x/y})\Delta\mu_{\textrm{mul},0,x/y},
\end{equation}
where $\Delta\mu_{\textrm{mul},0,x/y}=\exp\left(-\j\phi'_{\textrm{NL},0,x/y}\right)$ and $\Delta\mu'_{\textrm{add},0,x/y}=\Delta a'_{\textrm{ICIXPM},0,x/y}+\Delta a'_{\textrm{IFWM},0,x/y}$. It is important to note that in (\ref{eqn12}), $\phi'_{\textrm{NL},0,x/y}$, $\Delta a'_{\textrm{ICIXPM},0,x/y}$ and $\Delta a'_{\textrm{IFWM},0,x/y}$ are obtained by approximating transmitted symbols $a_{i,x/y}$ by $r_{i,x/y}$ in (\ref{eqn8}), (\ref{eqn9}), and (\ref{eqn10}).

\section{Learning Approaches for PB-NLC} 
\label{s:learned}
\begin{figure}[!t]
	\begin{centering}
\includegraphics[width=1.05\columnwidth]{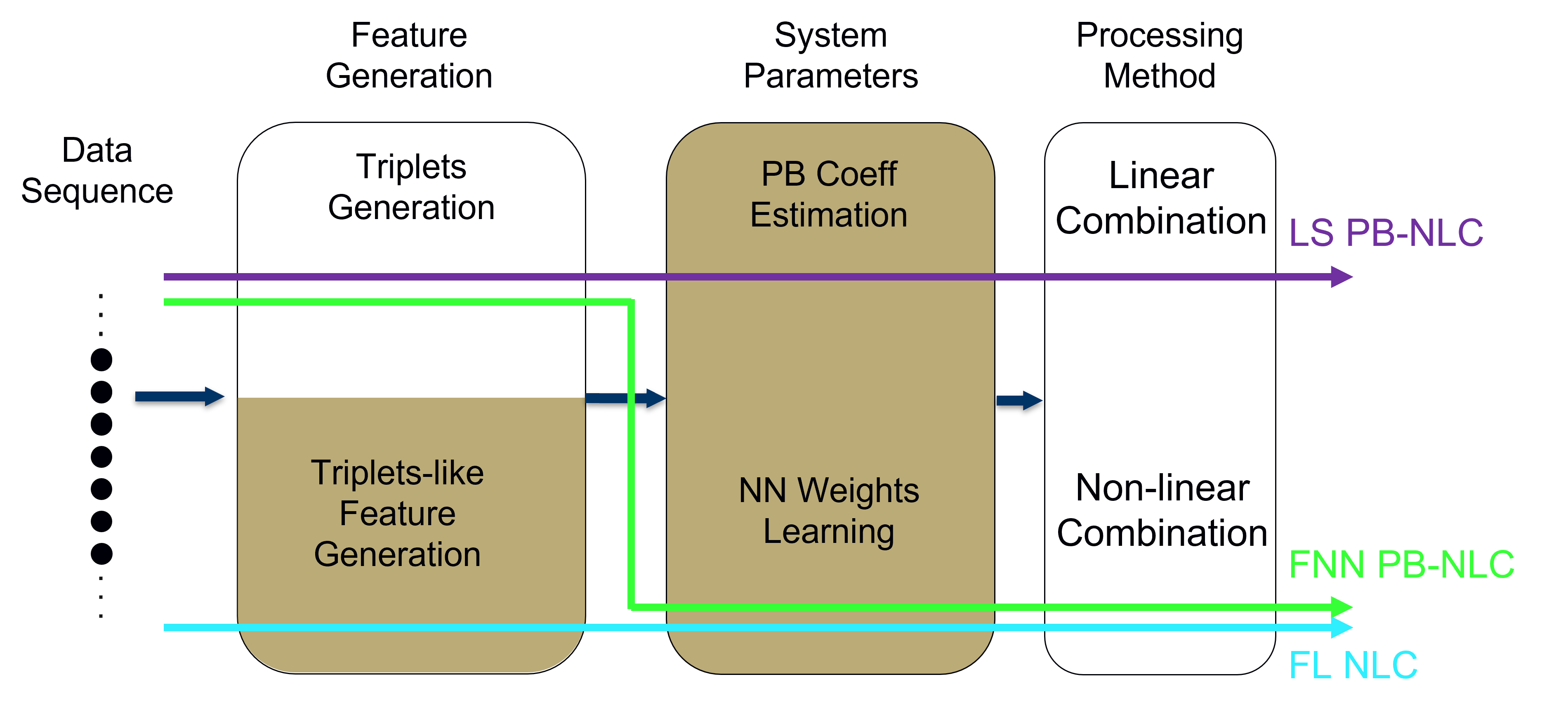}\caption{The schematic representation of existing learned/non-learned PB-NLC methods. 
The colored arrows show the steps the different methods take. The left box pertains to the generation of features, the center box to the learning of coefficients and weights, and the right box to the combining of features and coefficients/weights to estimate the distortion field estimation. The brown-shaded regions indicate learning-from-data steps.}
		\label{figstructure}
		\par\end{centering}
\end{figure}

A conceptual schematic diagram of the learned PB-NLC techniques is shown in Fig.~\ref{figstructure}.
In this section, we first briefly review the  traditional NN-based and  LS-based learned PB-NLC techniques, as well as present the new FNN-AM PB-NLC. These methods use  triplets of received data samples as the input features, which corresponds to the white-shaded part in left box in Fig.~\ref{figstructure}. 
We will then present our proposed FL NLC technique, where we learn the triplet-like features from the received samples. This path is shown in the brown-shaded region in the left box in Fig.~\ref{figstructure}. 

\subsection{Existing Learned PB-NLC Methods}
\subsubsection{LS PB-NLC}
Similar to the CONV PB-NLC summarized in Section~\ref{s:convpbnlc}, LS PB-NLC takes the triplets as the input and linearly combines them with the PB coefficients to generate the nonlinearity distortion field. However, instead of analytically computing the PB coefficients through numerical integration, the least squares method is applied to learn the PB coefficients, as shown in \cite{ARedyuk2020}.

\subsubsection{NN PB-NLC}
\label{s:nnpbnlc}
In the existing NN-based learned PB-NLC methods, the triplets computed from received samples are used as the input features. The distortion field is estimated by nonlinearly combining the features and the learned weights of the feature processing NN (see Fig.~\ref{figstructure}). FNN, RNN and CNN architectures have been explored in the literature \cite{SZhang2019, YZhao2020Access,Lietal}. Since there are little performance differences between those, and since previously mentioned differences in complexity are not substantial when considering complexity reduction methods for both feature generation and combining as done in this work, we adopt the FNN PB-NLC for our comparison in the following.

\subsection{The Proposed FNN-AM PB-NLC}
\label{s:AMPB}
\begin{figure}[t]
		\includegraphics[width=1\columnwidth,height=0.2\paperheight]{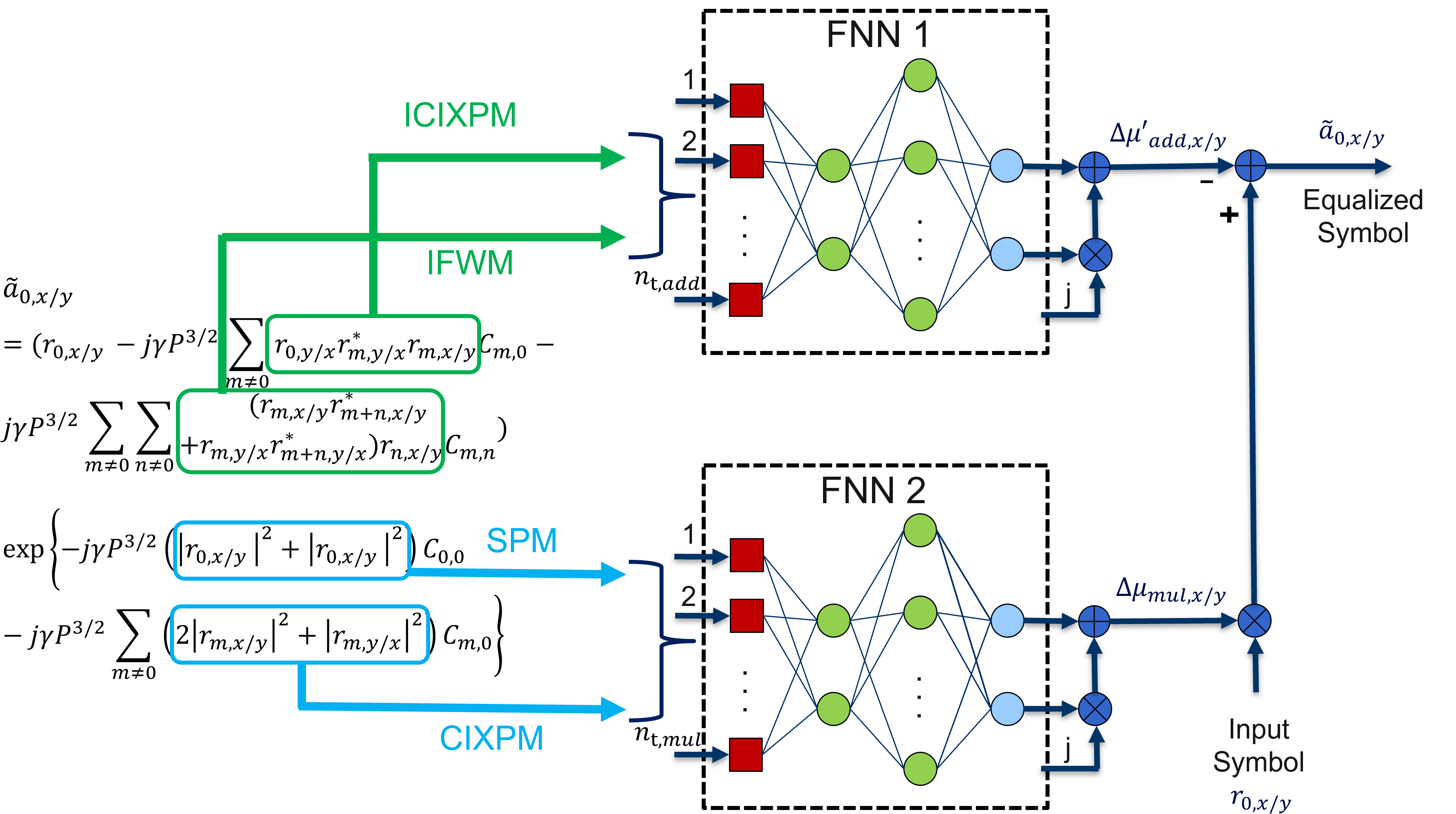}\caption{Illustration of the  FNN-AM PB-NLC technique. $M$ represents the number of SPM+CIXPM triplets, and $N$ is the number of ICIXPM+IFWM triplets (only the compensation of $x$-polarization data is shown for brevity).}
    \label{fig2}
\end{figure}

Our first extension to the existing methods is to apply  FNN triplets combining to the AM perturbation model, which we refer to as FNN-AM PB-NLC. The model structure is shown in Fig.~\ref{fig2}. The first FNN (FNN~1) is supplied with the triplets representing SPM and CIXPM, whereas the second FNN (FNN~2) processes  triplets for ICIXPM and IFWM as input. 
 The outputs of FNN~1 and FNN~2 are the real and imaginary parts of the $\Delta\mu_{\textrm{mul}}$ and $\Delta\mu'_{\textrm{add}}$, respectively. After estimating the multiplicative and additive parts of the nonlinear distortion, the NLC is performed according to (\ref{eqn12}).

\subsection{The Proposed FL NLC}
The PB-based learned NLC methods described above take the triplets as the input features to estimate the nonlinearity distortion field. However, the computational complexity to generate the triplets is considerably high. To address this issue, we propose to use a feature learning NN that replaces the triplet computation. We expect two potential benefits from feature learning. First, approximations underlying to first-order PB-NLC can be adjusted for in the feature generation. Second, the NN-based learning permits the application of complexity reduction through network pruning \cite{LTH}, which is more fine-grained than discarding triplets. 

\begin{figure}[t]
	\begin{centering}
	\includegraphics[width=1\columnwidth]{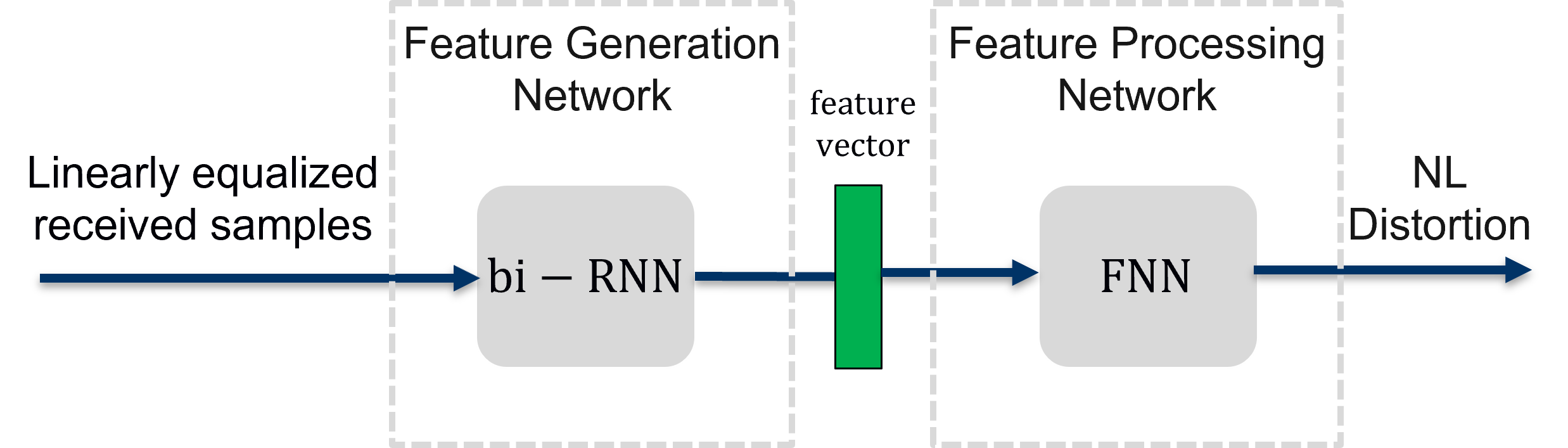}\caption{Illustration of the structure of the proposed FL NLC.}
		\label{fig6}
		\par\end{centering}
\end{figure} 
The principle structure of FL NLC is shown in Fig.~\ref{fig6}. It consists of a feature generation NN and a feature processing NN. For the latter, we adopt an FNN as used in the NN-based learned PB-NLC methods from Sections~\ref{s:nnpbnlc} and \ref{s:AMPB}. Our choice for the feature generation NN is a bidirectional RNN (bi-RNN). This is deemed meaningful since the triplets for estimating the distortion field for adjacent data symbols have many common elements, i.e., linearly equalized received samples. Therefore, the memory of a bi-RNN can be used to exploit this overlap to generate successive feature samples. While a basic RNN cell is simple, it faces the vanishing gradient problem \cite{Bengio}. Hence, we instead consider LSTM and GRU cells \cite{Chung2014}, i.e., bi-LSTM and bi-GRU for feature generation.   


\begin{figure*}[!t]
	\begin{centering}
	\includegraphics[width=0.8\textwidth,height=0.45\textheight]{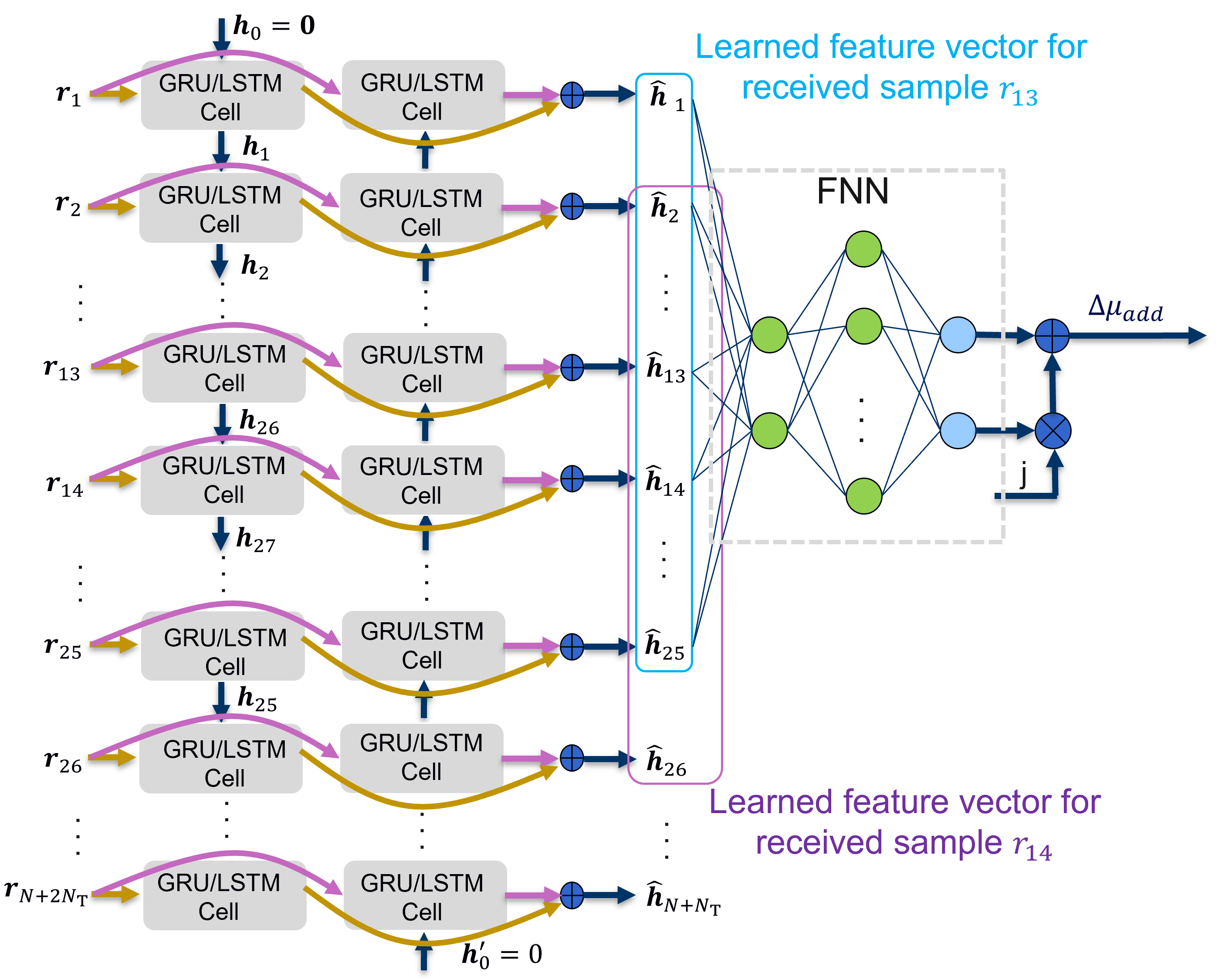}\caption{Feature generation network for the FL NLC technique. The feature processing FNN on the right is used for to estimate the nonlinearity distortion field. The blue and purple windows show examples for feature vectors when $\Nt=12$.}
		\label{fig7}
		\par\end{centering}
\end{figure*} 
Fig.~\ref{fig7} shows the detailed network structure to generate the learned features. The input samples to the network $\boldsymbol{r}_{i}$ contain the real and imaginary parts of linearly equalized received samples from both x- and y-polarization. 
In the feature generation part, one frame consists of $N+2\Nt$ samples to extract the features for $N$ symbols, i.e., the $2\Nt$ extra samples account for forward and backward transients in the memory cells. At each time step $i$, the GRU/LSTM cells take the  hidden state $\boldsymbol{h}_{i-1}$ ($\boldsymbol{h}'_{N+2\Nt-i}$ in the backward direction) and input $\boldsymbol{r}_{i}$ to generate the updated hidden state $\boldsymbol{h}_{i}$ ($\boldsymbol{h}'_{N+2\Nt-i+1}$ in the backward direction). The length of $\boldsymbol{h}_{i}$ and $\boldsymbol{h}'_{i}$ is $u$, where $u$ is the hidden unit of GRU/LSTM cells.
The updated forward and backward hidden states $\boldsymbol{h}_{i}$ and $\boldsymbol{h}'_{N+2\Nt-i+1}$ are then summed up to generate a feature stamp $\hat{\boldsymbol{h}}_{i}$. 
Then, the feature vector $\boldsymbol{F}_{i}$ for the received sample $\boldsymbol{r}_{i}$ is formed by concatenating adjacent feature stamps using a sliding window with length $2\Nt+1$, i.e., $\boldsymbol{F}_{i}= [\hat{\boldsymbol{h}}_{i-\Nt}, \ldots, \hat{\boldsymbol{h}}_{i}, \ldots, \hat{\boldsymbol{h}}_{i+\Nt}]$. As a result, the size of the feature vector for each received sample is $u(2\Nt+1)$. Lastly, this feature vector is input to the processing FNN to estimate the nonlinearity distortion field $d_{i,x/y}$ in \eqref{eqn2}.

We conclude by noting that in \cite{Ming2022}, a nonlinearity compensation approach using a bi-LSTM with $\Nt=1$ has been proposed. However, the bi-LSTM of \cite{Ming2022} is a black-box approach that is not related to a  nonlinearity model.  The proposed FL NLC follows from the PB nonlinearity model and a learned approach for feature generation and processing. This model-based NN design is in generally more potent than a black-box approach.

\section{Complexity Evaluation}
\label{s:complexity}

In this section, we evaluate the computational complexity of the presented perturbation-based methods in terms of real-valued multiplications per symbol, and we evaluate one complex-valued multiplications as 4 real-valued multiplications. We note that all PB-NLC based methods operate at symbol rate. We focus on the operational complexity, i.e., the computations involved in estimating the nonlinear distortion once the PB-NLC scheme has been set up or learned. First, we present the expressions for the baseline complexities which do not consider complexity reduction for triplet generation or triplet processing. We then discuss methods how complexity can be reduced.

\subsection{Baseline Complexity}
The triplets-based methods naturally require the computation of triplets. The straightforward approach to compute $\nt$ triplets involves $8\nt$ real-valued multiplications per symbol, as we use the term in the parentheses in \eqref{e:triplets} for both x- and y-polarization. Accounting for the additional multiplication with the complex-valued PB coefficients, the overall complexity for CONV PB-NLC is given by
\begin{equation}
C_{\mathrm{CONVPB}} = 12\nt.
\end{equation}

The CONV-AM PB-NLC rearranges the perturbation equation, resulting in a slightly different complexity than the CONV PB-NLC technique. The overall triplets generation complexity is composed of four independent parts corresponding to the IFWM, ICIXPM, CIXPM and SPM nonlinearity terms in \eqref{eqn7}, in which each component has a complexity of $8n_{\mathrm{IFWM}}$, $4n_{\mathrm{ICIXPM}}+4$, $2n_{\mathrm{CIXPM}}+2$ and $6$ real-valued multiplications per symbol, respectively. The variables $n_{\mathrm{IFWM}}$, $n_{\mathrm{ICIXPM}}$, and $n_{\mathrm{CIXPM}}$ are the number of triplets for the corresponding nonlinearity terms.
Hence, assuming that the exponential function is implemented as a look-up table, the overall number of real-valued multiplications per symbol is
\begin{equation}
C_{\mathrm{CONVAM}} = 12n_{\mathrm{IFWM}}+8n_{\mathrm{ICIXPM}}+4n_{\mathrm{CIXPM}}+ 2n_{\mathrm{SPM}}+12.
\end{equation}

For LS PB-NLC, there is one additional complex-valued multiplication on the received symbol (shown in equation (5) in \cite{ARedyuk2020}). Therefore, the complexity of LS PB-NLC is given by
\begin{equation}
C_{\mathrm{LS}} = 12\nt+4.
\end{equation}

For FNN PB-NLC, the number of multipliers in a NN is actually
the number of trainable weights present in all the layers. We assume a NN with $L$ layers, and $N_{\ell}$ neurons at layer $\ell$.  We note that $N_{0}=2\nt$ as the input features are the real and imaginary parts of the triplets. 
 The total number of real-valued multiplications is  
\begin{equation}
C_{\mathrm{FNN}} = 8\nt+\sum_{\ell=1}^{L}N_{(\ell-1)}N_{\ell}.
\end{equation}

For our proposed FNN-AM PB-NLC, we apply two FNNs to process the additive triplets and multiplicative triplets separately.
We can express the total complexity as 
\begin{multline}
C_{\mathrm{FNNAM}} = 8n_{\mathrm{IFWM}}+4n_{\mathrm{ICIXPM}}+2n_{\mathrm{CIXPM}}+2n_{\mathrm{SPM}}+12
\\
+\underbrace{\sum_{\ell_{1}=1}^{L_{1}}N_{(\ell_{1}-1)}N_{\ell_{1}}}_{\mbox{FNN1}}+\underbrace{\sum_{\ell_{2}=1}^{L_{2}}N_{(\ell_{2}-1)}N_{\ell_{2}}}_{\mbox{FNN2}}.
\end{multline}

Finally, for FL NLC, $N+2\Nt$ samples are fed into a bi-RNN with $u$ hidden units to generate a chain of features $\boldsymbol{h}_{t}$, to  compensate the center $N$ samples. 
If an LSTM is used, we consider the operations of a single LSTM cell described by
\begin{align}
\label{LSTMeqn}
\begin{split}
 \boldsymbol{i}_{t} &= \epsilon(\boldsymbol{W}_{i}\boldsymbol{h}_{t-1}+\boldsymbol{U}_{i}\boldsymbol{r}_{t}+b_{i}) ,
\\
 \boldsymbol{f}_{t} &= \epsilon(\boldsymbol{W}_{f}\boldsymbol{h}_{t-1}+\boldsymbol{U}_{f}\boldsymbol{r}_{t}+b_{f}) ,
\\
 \boldsymbol{o}_{t} &= \epsilon(\boldsymbol{W}_{o}\boldsymbol{h}_{t-1}+\boldsymbol{U}_{o}\boldsymbol{r}_{t}+b_{o}) ,
\\
 \boldsymbol{g}_{t} &= \tanh(\boldsymbol{W}_{g}\boldsymbol{h}_{t-1}+\boldsymbol{U}_{g}\boldsymbol{r}_{t}+b_{g}) ,
\\
 \boldsymbol{c}_{t} &= \epsilon(\boldsymbol{f}_{t}\odot \boldsymbol{c}_{t-1} + \boldsymbol{i}_{t}\odot \boldsymbol{g}_{t}) ,
\\
 \boldsymbol{h}_{t} &= \tanh(\boldsymbol{c}_{t})\odot \boldsymbol{o}_{t}, 
\end{split}
\end{align}
where $\odot$ denotes the point-wise multiplication of two vectors. The real-valued matrices $\boldsymbol{W}$ and $\boldsymbol{U}$ have a size of $u
times u$ and the real-valued vectors $\boldsymbol{i}_{t}, \boldsymbol{f}_{t}, \boldsymbol{o}_{t}$, $\boldsymbol{g}_{t}$ and $\boldsymbol{c}_{t}$ have a length of $u$. 
The efforts to compute each of $\boldsymbol{i}_{t}, \boldsymbol{f}_{t}, \boldsymbol{o}_{t}$ and $\boldsymbol{g}_{t}$ are $u(u+4)$, and to compute each of $\boldsymbol{c}_{t}$ and $\boldsymbol{h}_{t}$, we need $2u$ and $u$ multiplications, corespondingly. Therefore, for a single LSTM cell, the complexity is $4u(u+4)+3u$. Here we consider bidirectional LSTM consisting of two LSTMs, so the total complexity of Bi-LSTM is $8u(u+4)+6u$. Besides, we append $\Nt$ transient data at the beginning and end, so the total complexity should be multiplied by $(N+2\Nt)/N$. Then, together with the FNN complexity, the overall complexity of processing a single symbol by LSTM is
\begin{equation}
C_{\mathrm{FLLSTM}} = (8u(u+4)+6u)(N+2\Nt)/N+\sum_{\ell=1}^{L}N_{(\ell-1)}N_{\ell}.
\end{equation}
If a GRU is used, a similar computation can be applied. Considering the equations for a single GRU cell 
\begin{align}
\label{GRUeqn}
\begin{split}
 \boldsymbol{r}_{t} &= \epsilon(\boldsymbol{W}_{r}\boldsymbol{h}_{t-1}+\boldsymbol{U}_{r}\boldsymbol{r}_{t}+b_{r}) ,
\\
 \boldsymbol{z}_{t} &= \epsilon(\boldsymbol{W}_{z}\boldsymbol{h}_{t-1}+\boldsymbol{U}_{z}\boldsymbol{r}_{t}+b_{z}) ,
\\
 \boldsymbol{n}_{t} &= \tanh(\boldsymbol{W}_{n}(\boldsymbol{r}_{t}\odot \boldsymbol{h}_{t-1})+\boldsymbol{U}_{n}\boldsymbol{r}_{t}+b_{n}) ,
\\
 \boldsymbol{h}_{t} &= (\boldsymbol{1}-\boldsymbol{z}_{t})\odot \boldsymbol{n}_{t}+\boldsymbol{z}_{t}\odot \boldsymbol{h}_{t-1} ,
\end{split}
\end{align}
the total complexity to process a single symbol by GRU becomes 
\begin{equation}
C_{\mathrm{FLGRU}} = (6u(u+4)+6u)(N+2\Nt)/N+\sum_{\ell=1}^{L}N_{(\ell-1)}N_{\ell}.
\end{equation}

\subsection{Complexity Reduction Methods}
\label{CReduce}
\subsubsection{Triplets-based Methods}
Most of the complexity for all triplets-based methods comes from the triplet generation. Since some computations are repeatedly executed during triplets generation, a cyclic buffer (CB) to store and reuse intermediate results has been proposed in \cite{ARedyuk2020}. This store-and-reuse approach can significantly reduce the required number of computations as we will show in the numerical results in Section~\ref{s:results}.

For the triplets processing complexity, we first consider the symmetry of perturbation coefficients. In particular, it follows from equation~\eqref{eqn3} that the coefficients have the symmetry $C_{m,n}=C_{n,m}$. 
Hence, the processing complexity can be reduced nearly by half \cite{ZTao2013}. While this symmetry does not necessarily hold for the learned PB-NLC methods, we have observed only a negligible performance loss when imposing it.  

Furthermore, \cite{ARedyuk2020} proposed a quantization method using the K-means algorithm to cluster the coefficients in a complex plane, which reduces the triplet processing complexity with a potentially small performance degradation. This quantization can be applied for the linear triplets combination systems, i.e., LS PB-NLC and CONV(-AM) PB-NLC. For the NNs used in FNN(-AM) PB-NLC, we adopt the weight pruning method as in \cite{SZhang2019}, but instead of applying a predefined threshold, we gradually prune the relatively small weights in the first layer of the FNN, 
since it contributes most to the complexity. Then we fine-tune the rest of the network to compensate for the pruning loss. 

\subsubsection{FL NLC}
The complexity of the bi-RNN for feature generation shown in Fig.~\ref{fig7} is reduced by applying the same weight pruning method as in the triplets-based learned approaches mentioned above to the weights in GRU/LSTM. 
To reduce the complexity of the feature processing FNN, we found the K-means quantization to be more effective. We thus apply this quantization to the weights in the first layer of the FNN. 

\section{Numerical Results and Discussion}
\label{s:results}

In this section, we compare the performance and complexity of all PB-NLC-based methods. \footnote{The source code is available at https://zenodo.org/record/7933021\#.ZGLt83bMKM9}
Firstly, the performance will be evaluated and compared without any complexity reduction. Then, we apply the  methods discussed in Section \ref{CReduce}  and present the performance-complexity trade-offs for the different methods.  


\subsection{Simulation Setup}

We simulate a polarization-multiplexed 5-channel wavelength-division multiplexed (WDM) transmission at 32~Gbaud per channel. The modulation format considered is 16-quadrature amplitude modulation (QAM). 
As shown in Fig.~\ref{fig1}, the transmitter consists of a root-raised cosine (RRC) pulse shaper, a 50\% chromatic dispersion pre-compensation (pre-CDC) \cite{SZhang2019}, a PBC, and a booster EDFA to set the transmit signal power. The long-haul optical fiber transmission link consists of 10 spans of standard single-mode fiber (SSMF) with a span length of 100~km, a dispersion parameter of 17~ps.nm$^{-1}$.km$^{-1}$, a nonlinear parameter of 1.2~W$^{-1}$.km$^{-1}$, an attenuation coefficient of 0.2~dB.km$^{-1}$ and a PMD parameter of 0.1~ps.km$^{-\frac{1}{2}}$. An in-line EDFA with a 6~dB noise figure is used after each span to compensate for the signal power attenuation. 
In the receiver, the signal passes through a 50\% post-CDC, and a matched filter. Following that, a least mean square (LMS) adaptive 2x2 filter is used to compensate for PMD, and blind phase search (BPS) is applied for CPR. Then one of the concerned techniques is applied for nonlinearity compensation. Finally, the signal is demodulated to evaluate the performance. 

\paragraph{Triplets Selection}
For all the triplets-based methods, the number of triplets determines the overall complexity, so removing unimportant triplets can efficiently reduce the complexity. We adopt the triplet selection method as in \cite{SZhang2019}. Specifically, the triplet $F_{(m,n),x/y}$ is selected, when 
\begin{equation}
\label{e:trunc}
|n| \le \min\left\{\frac{\rho\lceil \Nw\rceil}{|m|}, \lceil \Nw\rceil\right\},
\end{equation}
where $\Nw$, as the window size of the triplets, determines the largest value of $m$ and $n$, and $\rho$ is a hyper-parameter limiting the maximum production of $m$ and $n$.

\paragraph{Network Hyper-parameters}
The feature processing FNNs used in FNN PB-NLC, FNN-AM PB-NLC, and FL NLC use the same architecture as in \cite{SZhang2019}. Accordingly, the FNNs consists of two hidden layers with two neurons in the first and ten neurons in the second hidden layer, and two neurons for the output layer. In each hidden layer, leaky ReLU  with a factor of 0.5 is used as the nonlinear activation function, and a dropout layer with a rate of 0.2 is applied after the second hidden layer to avoid overfitting. 

For FL NLC, the hidden units of the GRU or LSTM cells varies between 20 and 30 to investigate the performance-complexity trade-off. The sliding window parameter is selected as $\Nt=12$. 

\paragraph{Network Training}
The training for FNN-based NNs uses the \emph{Adam} optimizer with a learning rate of $10^{-3}$ and batch size of 100. The mean-squared error (MSE) is chosen as the loss function. 

To train FL NLC, we set $N=1$ and pad $N_{\mathrm q}=100$ symbols on each side for the transient effect, in addition to the $2\Nt$ symbols for the feature generation. Therefore, the input for the training is $\boldsymbol{r}_{i,\textrm{train}}=(\boldsymbol{r}_{i-\Nt-N_{\mathrm q} }, ..., \boldsymbol{r}_{i}, ...,\boldsymbol{r}_{i+\Nt+N_{\mathrm q}})$. Then the feature generated from the bi-RNN is $\boldsymbol{f}_{i}=(\hat{\boldsymbol{h}}_{i-\Nt}, ..., \hat{\boldsymbol{h}}_{i}, ..., \hat{\boldsymbol{h}}_{i+\Nt})$. 
To test FL NLC, we again use $N=2^{16}$ symbols. Furthermore, the \emph{Adam} optimizer with a learning rate of $10^{-4}$ is used for training and the batch size is 512. During training, we noticed that the MSE loss may continue to decrease when the resulting Q-factor performance does not improve anymore. In fact, the Q-factor performance may deteriorate with more training. Hence, we apply a form of early stopping based on the Q-factor performance.\footnote{We note that the reason for this behaviour manifests itself in a grid-shaped noise cloud around QAM symbols that occurs with longer training. A similar observation has recently been reported in \cite{Diedolo} when the MSE loss function is used to train an FNN-based nonlinear equalizers with received samples as input features, as it is the case for FL NLC. The authors proposed an entropy-regularized MSE loss function to rectify their distortion problem, which may also be an effective alternative for training FL NLC}.

We use $2^{18}$ training and $2^{16}$ test samples for all NN-based methods.

\begin{figure}[t]
	\begin{centering}
	\includegraphics[width=1\columnwidth]{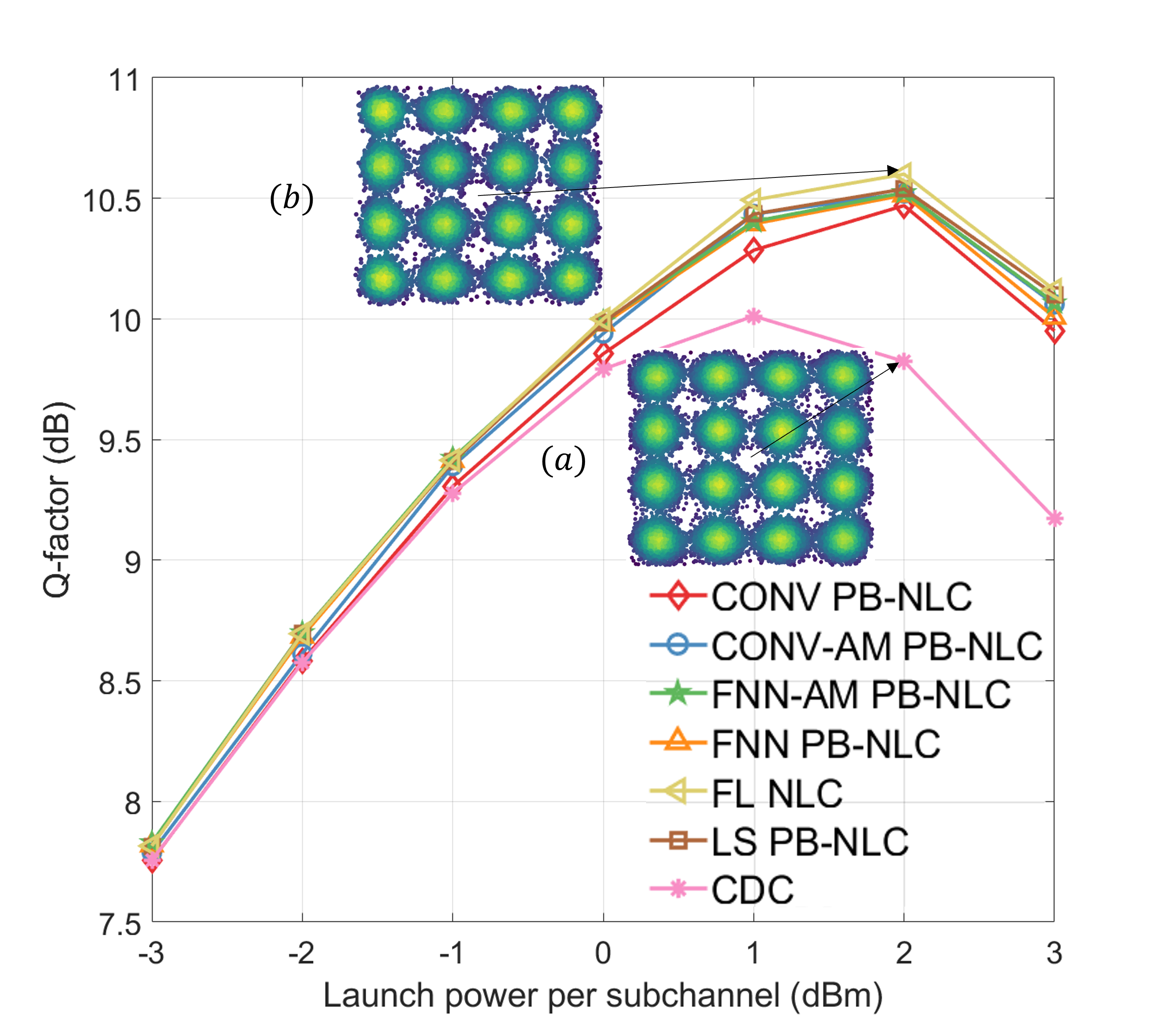}\caption{Performance comparison for learned and non-learned PB-NLC methods. Inset (a) and (b) show the received constellations before and after applying FL NLC at 2~dBm launch power, respectively.}
		\label{figPerformance}
		\par\end{centering}
\end{figure} 
\begin{figure*}[!t]
	\begin{centering}
	\includegraphics[width=0.95\textwidth,height=0.53\textheight]{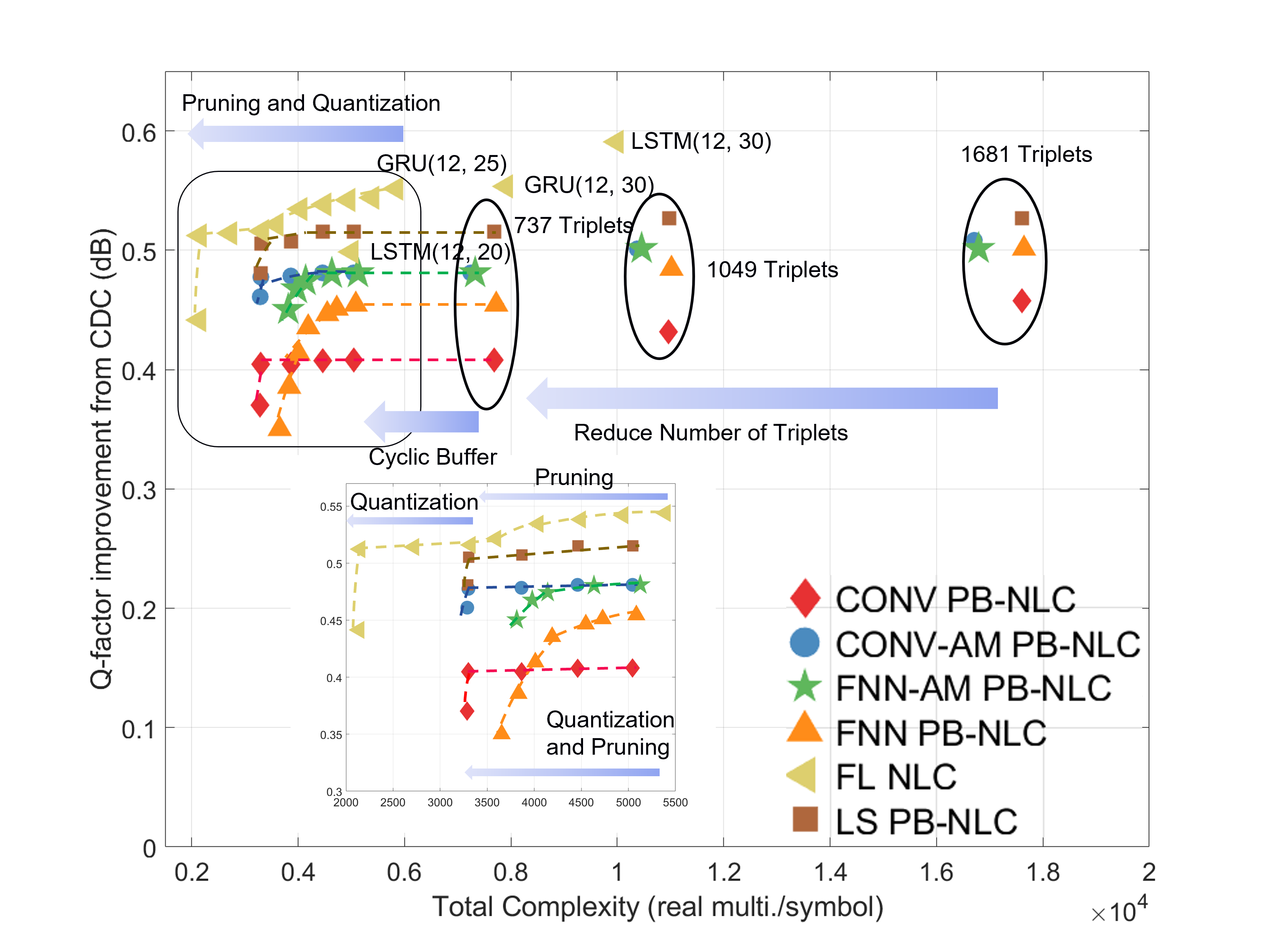}\caption{Performance-complexity chart for learned and non-learned PB-NLC methods. The effects of the applied complexity reduction methods are identified as arrows. For FL NLC, GRU/LSTM($u,\Nt$) means bi-GRU/LSTM with $u$ hidden units and sliding-window parameter $\Nt$.}
		\label{figPC}
		\par\end{centering}
\end{figure*}

\subsection{Performance Results}
For all results, the center WDM channel is chosen as the target channel to evaluate the performance. 

\subsubsection{Q-factor performance} First, we consider the error-rate performance when no complexity reduction methods are applied. For the triplets-based techniques, we start with a sufficiently large number of triplet features to assess the maximal performance. In particular, the window size $\Nw=75$ is selected, and the number of triplets obtained after applying the truncation criterion in \eqref{e:trunc}is $\nt=1681$, and $n_{\mathrm{IFWM}}=1378$, $n_{\mathrm{ICIXPM}}=151$, $n_{\mathrm{CIXPM}}=151$, and $n_{\mathrm{SPM}}=1$. Similarly, for FL NLC, a bi-LSTM with $u=30$ hidden units is used. 

Fig.~\ref{figPerformance} shows the Q-factor performance for the considered PB-NLC techniques as well as for CDC-only processing as a reference. The inset (a) and (b) in Fig.~\ref{figPerformance} show the received constellations before and after applying FL NLC at 2~dBm launch power, respectively. We observe that all learned triplets-based methods perform fairly similarly and achieve essentially the same optimal Q-factor at a transmission power of 2~dBm per channel. Besides, the CONV-AM PB-NLC reaches an almost identical optimal Q-factor. Hence, without restrictions on complexity, there is little benefit to the learned approaches. Furthermore, our proposed FL NLC performs slightly better than the triplets-based methods and shows a cleaner constellation than CDC. 

\subsubsection{Performance-Complexity Comparison}
\label{PCC}
Next, we evaluate the peak Q-factor performance when reducing the computational complexity for the different PB-NLC techniques. 

Fig.~\ref{figPC} shows the Q-factor improvement over CDC as a function of the computational complexity measured as the number of real-valued multiplications per symbol. 
For the triplets-based methods, the complexity reduction is achieved first through smaller symbol-window sizes, which, when shortened from $\Nw=75$ to $\Nw=37$ reduces the number of triplets from $\nt=1681$ to $\nt=737$, which includes $n_{\mathrm{IFWM}}=586$, $n_{\mathrm{ICIXPM}}=75$, $n_{\mathrm{CIXPM}}=75$, and $n_{\mathrm{SPM}}=1$.  
Since the performances of CONV-AM PB-NLC, FNN-AM PB-NLC and LS PB-NLC remain close to the best possible values with just an onset of degradation, we fix $\nt=737$ triplets and next further reduce complexity by making use of CB. This computationally efficient way of generating triplets does not degrade performance, as seen in Fig.~\ref{figPC}. 
Finally, the number of computations required for processing triplets is lowered through weight pruning for the NNs and coefficient quantization using the K-means quantization for the other triplets-based methods, respectively.

For FL NLC, we first compare bi-LSTMs and bi-GRUs with different hidden units, as shown in Fig.~\ref{figPC}. Since the performance and complexity of the bi-GRU with 25 hidden units are comparable to those of the triplets-based methods, we start the weight pruning at this point. We stop the pruning process before a significant performance degradation, and then we further reduce the processing complexity by the K-means quantization. 

The result details are highlighted in the inset in Fig.~\ref{figPC}.
We observe that for all the triplets-based methods, LS PB-NLC achieves the best performance-complexity trade-off, followed by CONV-AM PB-NLC. This suggests that PB-NLC based on models that process triplets linearly together with coefficient quantization are more effective than nonlinear processing with FNNs and weight pruning. 
Furthermore, our proposed FL NLC performs better than all of the triplets-based methods. In particular, we observe that it compares favourably with the other methods when considering only pruning for complexity reduction. Finally, the complexity of FL NLC solution is further reduced through quantization of the feature processing FNN. This provides about 35\% complexity reduction compared to LS PB-NLC with similar performance. 

\section{Conclusions}
\label{s:conclusions}

In this paper, we have conducted a comprehensive complexity-performance trade-off analysis of the learned and non-learned PB-NLC methods existing in the literature that considers the noiseless Manakov equation as the channel model. In doing so, we have also introduced the FNN-AM PB-NLC extension to the learned PB-NLC methods, and we have proposed the FL NLC technique that uses a bi-LSTM/bi-GRU network for learning triplet-like input features and an FNN for the feature processing. Our numerical results indicate that among the triplets-based methods, the LS PB-NLC approach that linearly combines triplets is more effective than the nonlinear combining using FNNs.
Furthermore, the new FL NLC technique outperforms all the other learned and non-learned PB-NLC methods. Thus, the combination of feature learning and learned processing seems to be the most promising approach in the context of PB-based compensation for the fiber nonlinearity effect in a long-haul optical fiber link. An interesting question for further studies is an alternative training scheme for FL NLC which does not exhibit the grid-shaped constellation-distortion phenomenon.

\ifCLASSOPTIONcaptionsoff
  \newpage
\fi



%
\bibliographystyle{IEEEtran}
\bibliography{Reference}
	
	

%








\end{document}